# Catalytic precursor dissociation in Hot-Wire CVD and comparing a-Si:H growth under continuous and pulsed silane flow conditions

# Comparing Hot-Wire CVD growth of a-Si:H under continuous and pulsed silane flow conditions Catalytic precursor dissociation model

Swati Goyal[1], Karam Veer Singh[1], Rajiv O. Dusane[1] and Triratna P. Muneshwar[1,c]

[1]*Department of Metallurgical Engineering and Materials Science, Indian Institute of Technology Bombay, Mumbai, 400076, India*

[c]*Corresponding Author*

Hot-wire Chemical vapor deposition (HWCVD) of hydrogenated amorphous silicon (a-Si:H) thin films utilizes the dissociation of silane ($SiH_4$) precursor over heated tungsten or tantalum filaments (≥ 1600 °C). In this work, assuming catalytic dissociation mechanism, we present kinetic modelling of this decomposition reaction and the resulting a-Si:H film growth. Our model calculations showed that for an identical dose of the introduced $SiH_4$ precursor, a-Si:H thickness was considerably higher for the pulsed $SiH_4$ flow as compared to the continuous $SiH_4$ flow. The pulsed $SiH_4$ flow is represented by time intervals $t_{ON}$ and $t_{OFF}$, where the $SiH_4$ flow rate ($F_{SiH_4}$) is at the set-point and zero, respectively. In agreement with our model calculations for an introduced 75 $cm^3$ (STP) $SiH_4$ dose, the resulting a-Si:H film thickness was 175 ± 5 nm under continuous precursor flow, whereas it considerably increased to 425 ± 8 nm when this $SiH_4$ dose was split into 15 shorter pulses ($t_{ON}$ =15s; $t_{OFF}$ = 60s). Moreover, these a-Si:H films deposited using pulsed $SiH_4$ flow exhibited improved electrical properties, with a dark conductivity ($\sigma_d$) of $1.1\times10^{-11}$ S/cm and a photoconductivity ($\sigma_{ph}$) of $\sim5.8\times10^{-5}$ S/cm, compared to films deposited under continuous $SiH_4$ flow ($\sigma_d$ $\sim2.5\times10^{-10}$ S/cm and $\sigma_{ph}$ $\sim3.5\times10^{-6}$ S/cm).

## I. INTRODUCTION:

Hydrogenated amorphous silicon (a-Si:H) is a material of considerable interest due to its optoelectronic properties and low processing temperature for integration into a variety of thin film device applications such as photovoltaics, thin film transistors and sensors.[1–3] Incorporated H in a-Si:H is effective in passivating the dangling/unsaturated bonds in the amorphous network, which reduces bulk defect density, thereby improving electronic charge transport in the a-Si:H.[4] Over the years, Hot-Wire Chemical Vapor Deposition (HWCVD, also known as catalytic CVD or cat-CVD) has emerged as a promising alternative to the Plasma Enhanced CVD (PECVD) for the growth of high-quality a-Si:H films.[5,6] In the HWCVD process, silane ($SiH_4$) precursor flowing over resistively heated filaments (Ta or W, heated to ≥1600 °C), undergoes decomposition to generate reactive species (Si, $SiH_x$ where x = 1, 2, or 3, and atomic H) which contribute to the a-Si:H growth.[7,8] Since these heated filaments are physically separated from the substrate by a few centimeters, a-Si:H film growth could be realized at lower substrate temperatures in HWCVD as compared to PECVD.[9] Also the HWCVD reactor configuration is relatively straightforward,[10] utilizes $SiH_4$ precursor more efficiently,[6] and is readily adaptable for growth over the large area substrates.[11] Furthermore, the electronic characteristics of the HWCVD grown a-Si:H films are comparable with those obtained with PECVD.[12,13] However, regardless of all the above-mentioned advantages, HWCVD adoption has been limited, primarily due to the filament ageing-related drift in growth rate,[14,15] which require periodic replacement of W (or Ta) filaments.[16]

In this article, we propose a phenomenological model of precursor reaction over a hot filament during HWCVD growth, assuming a catalytic dissociation exclusively taking place at the active sites $M^*$ on the filament surface as shown schematically in Fig. 1(a). For $SiH_4$ dissociation, we considered that the impinging $SiH_4$ molecule must adsorbs on the surface site $M^*$ as $SiH_4^*$ for dissociation into $Si^*$ and $H^*$, which later desorb as Si and $H_2$ from the filament (* denotes surface species). The progress of $SiH_4$ dissociation is tracked with the changing concentrations of the surface species ($M^*$, $SiH_4^*$, $Si^*$, and $H^*$) using model parameters $k_{in}$ for the normalized impingement rate of the $SiH_4$ molecule, $k_{d1}$ for desorption rate of $SiH_4^*$, $k_r$ for dissociation rate of $SiH_4^*$, $k_{d2}$ for desorption rate of $Si^*$ as Si(g) and $k_{d3}$ for desorption rate of $H^*$ as $H_2$(g). Effect of $SiH_4$ partial pressure $P_0$ is accounted within $k_{in}$, while the effect of filament temperature $T_f$ could be modelled considering temperature-dependent dissociation and desorption rate constants. In our catalytic dissociation model, only those $SiH_4$ molecules with access to $M^*$ sites can undergo dissociation and thereby contribute to film growth. Other $SiH_4$ molecules are assumed to remain un-dissociated and exit reactor unutilized. At any instant during precursor dissociation, the active site $M^*$ on the filament surface are consumed with $SiH_4$ adsorption, and later re-generated with desorption of every $SiH_4^*$, $Si^*$, and $H^*$ surface units. In a conventional HWCVD process with continuous precursor flow, the steady state concentration of the accessible active sites $[M^*]^{eq}$ is significantly lower than its initial concentration $C_0(M^*)$ because the active sites occupied by surface species ($SiH_4^*$, $Si^*$, and) are not accessible for precursor reaction.

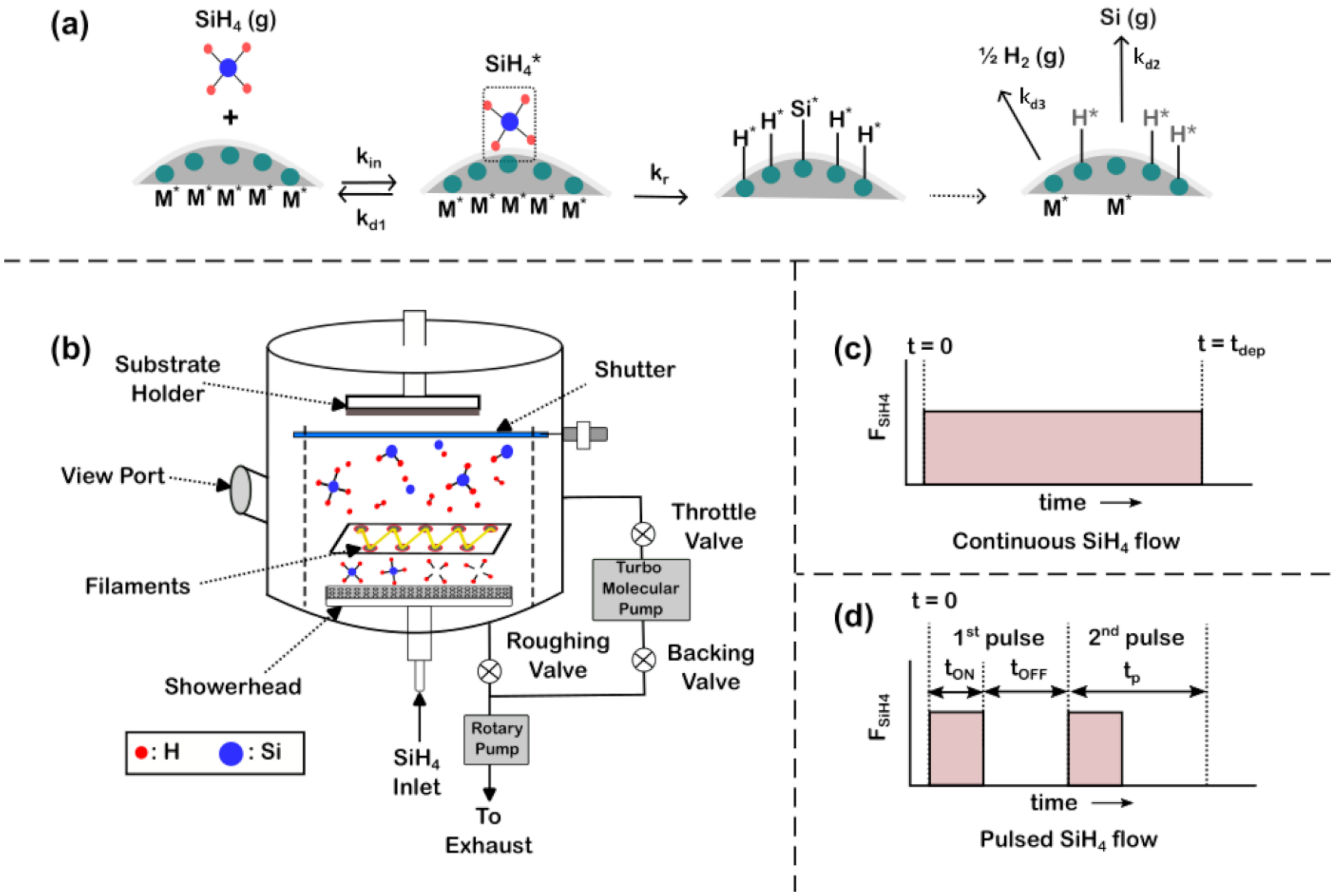

Fig. 1 Schematic representation of (a) proposed catalytic dissociation of $SiH_4$ on heated metal (Ta or W) filament during a-Si:H HWCVD process, (b) Configuration of HWCVD reactor used for our depositions, and representation of time-dependent $SiH_4$ flow ($F_{SiH_4}$) during HWCVD growth under (c) continuous and (d) pulsed precursor flow conditions.

Drawing parallels between the self-limiting ALD surface reactions[17] and our proposed catalytic precursor dissociation model, we hypothesized that pulsed precursor flow should lead to thicker films for an identical precursor dose (quantified as cubic centimeter at STP for $SiH_4$). Here, the pulsed $SiH_4$ flow is characterized with intervals $t_{ON}$ and $t_{OFF}$, where precursor flow is at the set-point or zero respectively. From calculations, we studied the effect of pulsed $SiH_4$ flow on the model a-Si:H growth and compared the results with those for continuous $SiH_4$ flow. These results are compared with a-Si:H HWCVD growth experiments under continuous and pulsed $SiH_4$ flow conditions. Using the dark conductivity ($\sigma_d$) and the photoconductivity ($\sigma_{ph}$) measurements, the resulting films were compared to highlight the effect of precursor pulsing on the overall a-Si:H film properties.

## II. PRECURSOR DISSOCIATION MODEL

In our proposed model, we assume that the precursor dissociation on the heated filament in the HWCVD reactor is catalytic in nature with a negligible contribution from thermal cracking. Here, we consider that the catalytically active sites $M^*$ are uniformly distributed on the filament surface, with their initial surface concentration $C_0(M^*)$ which depends upon the filament history,[18] its surface roughness,[18] and may further depend upon filament temperature.[19,20] In case of $SiH_4$ dissociation during HWCVD growth of a-Si:H films, the impinging rate of $SiH_4$ molecules on heated filament surface depends upon precursor flow rate $\left(F_{SiH_4}\right)$ and its partial pressure $\left(P_{SiH_4}\right)$, total filament area $(A_f)$ and filament configuration as described in Fig. 1(b). In our proposed model, all these factors are lumped together into an effective impingement rate $k_{in}$ and a more elaborate model accounting for each individually may be presented in the future.

For dissociation, the gas-phase $SiH_4$ molecule arriving at the filament surface must have a direct access to the surface reaction site $M^*$, otherwise it is pumped out of the reactor unutilized. Upon arrival, the $SiH_4$ molecule binds with this $M^*$ site to form an intermediate surface species $SiH_4^*$, which may either desorb as $SiH_4$ gas, freeing up the occupied $M^*$ site (see Fig. 1 (a) and R1) or undergo further dissociation. For this $SiH_4^*$ to dissociate into a $Si^*$ and 4 $H^*$ units in our proposed model, it must have access to additional 4 $M^*$ sites in its vicinity (see Fig. 1 (a) and R2). Desorption of $Si^*$ and $H^*$ from filament surface generates $Si(g)$ and $H_2(g)$ units which are considered to result in a-Si:H film growth on the substrate (see Fig. 1 (a), R3, R4).

$$SiH_4(\mathrm{g}) + M^* \underset{k_{d1}}{\overset{k_{in}}{\rightleftarrows}} SiH_4^* \qquad \text{(R1)}$$

$$SiH_4^* + 4M^* \xrightarrow{k_r} Si^* + 4 \cdot H^* \qquad \text{(R2)}$$

$$Si^* \xrightarrow{k_{d2}} Si(\mathrm{g}) + M^* \qquad \text{(R3)}$$

$$H^* \xrightarrow{k_{d3}} \frac{1}{2} H_2(\mathrm{g}) + M^* \qquad \text{(R4)}$$

Assuming elementary surface reactions, the kinetic expressions for reactions (R1 – R4) are represented by equations Eq. 1 – Eq. 5. Here the impingement rate $k_{in}$ and the reaction rate constants ($k_r$, $k_{d1}$, $k_{d2}$ and $k_{d3}$) are the model parameters; $[SiH_4^*]$, $[Si^*]$, $[H^*]$ and $[M^*]$ are the concentrations of $SiH_4^*$ , $Si^*$, $H^*$ and $M^*$ surface units normalized with respect to $C_0(M^*)$; $N(Si)$ and $N(H_2)$ denotes the total number of gaseous species $Si(g)$ and $H_2(g)$ evolved per unit area of the filament and normalized with respect to the $C_0(M^*)$. Normalized surface concentrations ($[SiH_4^*]$, $[Si^*]$, $[H^*]$ and $[M^*]$) can only vary between 0 and 1, and conservation of $M^*$ sites require that at any instance $[M^*]$ is given by Eq. 6.

$$\frac{d[SiH_4^*]}{dt} = k_{in}[M^*] - k_{d1}[SiH_4^*] - k_r[SiH_4^*][M^*]^4 \qquad \text{(Eq. 1)}$$

$$\frac{d[Si^*]}{dt} = k_r[SiH_4^*][M^*]^4 - k_{d2}[Si^*] \qquad \text{(Eq. 2)}$$

$$\frac{d[H^*]}{dt} = 4 \cdot k_r [SiH_4^*][M^*]^4 - k_{d3}[H^*] \quad \text{(Eq. 3)}$$

$$\frac{dN(Si)}{dt} = k_{d2}[Si^*] \quad \text{(Eq. 4)}$$

$$\frac{dN(H_2)}{dt} = \frac{1}{2} \cdot k_{d3}[H^*] \quad \text{(Eq. 5)}$$

$$[M^*] = 1 - ([SiH_4^*] + [Si^*] + [H^*]) \quad \text{(Eq. 6)}$$

The effect of filament temperature ($T_f$) on the overall dissociation kinetics is considered by assuming the Arrhenius expression for rate constants. Before the introduction of $SiH_4$ flow, the pristine hot filament surface is assumed to be devoid of $SiH_4^*$, $Si^*$ and $H^*$ surface species. With the initial condition: $[SiH_4^*] = 0$, $[Si^*] = 0$, $[H^*] = 0$, and $[M^*] = 1$; the precursor dissociation during conventional HWCVD growth is obtained from solving equations Eq. 1 – Eq. 5 with a constant $k_{in}$ throughout deposition as shown in Fig. 1 (c). We further assume that for short depositions lasting ~ 20-30 minutes, the diffusion of the surface species into the filament bulk is negligible. To simulate $SiH_4$ dissociation for pulsed precursor flow, the impingement rate $k_{in}$ is modulated over the pulse duration ($t_{ON} + t_{OFF}$) as $k_{in} = k_{in}^{set}$ for $0 \leq t < t_{ON}$ and $k_{in} = 0$ for $t_{ON} < t \leq (t_{ON} + t_{OFF})$ as shown in Fig. 1 (d). Here, we have assumed that the impingement rate $k_{in}$ follows an ideal rectangular waveform with abrupt switching between $k_{in}^{set}$ and 0 at $t = t_{ON}$ and $t = t_{ON} + t_{OFF}$ respectively.

In this study, we have limited our discussion to $SiH_4$ dissociation on the filament surface during HWCVD process, and did not explicitly model a-Si:H film growth on the substrate. A detailed film growth model accounting for additional parameters (e.g. HWCVD reactor configuration, filament temperature, filament/substrate area, filament-to-substrate separation, mass transport, substrate reactions, etc), is beyond the scope of this reported study and may be reported in the future. Instead, we assume that the modelled growth rate ($G^m$) is proportional to the rate at which $Si(g)$ species evolve from the filament surface, i.e. $G^m = \varphi \cdot \frac{dN(Si)}{dt}$, where $\varphi$ is the proportionality constant. With our definition of model growth rate, the resulting film thickness ($d_f^m$) is given by $d_f^m = \int_0^{t_{dep}} G^m \cdot dt$, where $t_{dep}$ is the deposition time.

## III. EXPERIMENTAL METHODS AND MATERIAL CHARACTERIZATION

Hydrogenated amorphous silicon (a-Si:H) films were deposited in a home-built multi-chamber HWCVD system that has been described in earlier publications.[21] Within this multi-chamber system, a-Si:H films were deposited using high-purity silane (99.999% $SiH_4$ from Matheson Tri-Gas Inc.) in a chamber dedicated for intrinsic Si growth, as shown in Fig. 1 (b). Tantalum wire (99.95% purity, from Kurt J. Lesker) was 0.5 mm in diameter with 70 cm effective length and separation of ~ 7 cm from the substrate.

Substrate was transferred from the load-lock (≤ $10^{-2}$ Torr) into HWCVD chamber at base pressure of ≤ $10^{-5}$ Torr through an intermediate central wafer handler (≤ $10^{-6}$ Torr). Both the substrate and Ta filaments were resistively heated to temperature set point 200 °C and 1650 °C, respectively. Substrate temperature was monitored using embedded thermocouples and controlled within ±2 °C of the set-point using PID controllers. For heated Ta filaments, temperature was monitored using an optical pyrometer (Model: STMRISBSFV, Stek) and controlled within ±10 °C by means of adjusting input power. Using mass flow controllers (MKS Instruments), a 20 sccm $SiH_4$ flow was introduced into the chamber and the chamber pressure was maintained at $20\times10^{-3}$ Torr using a throttle butterfly valve.

In the conventional HWCVD process, a constant $SiH_4$ flow rate ($F_{SiH_4}$, in sccm) is maintained throughout the deposition time ($t_{dep}$) as illustrated in Fig. 1 (c). The amount of silane introduced for deposition could be represented as volume ($cm^3$) at STP conditions given by $F_{SiH_4} \times t_{dep}$ (STP: 0 °C and $1.01\times10^5$ Pa). However, for HWCVD growth with pulsed $SiH_4$ flow, as shown in Fig. 1 (c), the total $SiH_4$ dose is given by $F_{SiH_4} \times t_{ON} \times N_p$ and the net deposition time is $N_p \times (t_{ON} + t_{OFF})$, where $N_p$ is the number of pulses. Hence one may consider conventional HWCVD growth as a limiting case of HWCVD with pulsed precursor flow where $t_{ON} = t_{dep}$, $t_{OFF} = 0$, and $N_p = 1$.

a-Si:H films were deposited on Corning 7059 glass substrates (~20 mm × 20 mm). For growth characterization, film thicknesses ($d_f$) was obtained by averaging thicknesses measured at multiple locations using a Veeco, Dektak 150 stylus profilometer. For conventional HWCVD a-Si:H process, the growth rate was derived from the ratio of measured film thickness ($d_f$) and deposition time ($t_{dep}$). However, for HWCVD a-Si:H growth with pulsed $SiH_4$ flow, growth per pulse (GPP) was obtained from the ratio of the measured film thickness ($d_f$) and the number of precursor pulses ($N_p$). To compare HWCVD growth under continuous and pulsed precursor flow, we further define process metrics (i) process throughput ($\zeta$) given by the ratio $\zeta = \frac{GPP}{(t_{ON}+t_{OFF})}$ and (ii) precursor utilization ($\eta$) that is proportional to $\eta \propto \frac{GPP}{t_{ON}}$ that represents how efficiently the introduced precursor is utilized towards film growth. Effect of $SiH_4$ pulse parameters ($t_{ON}$ and $t_{OFF}$) on growth characteristics (GPP, $\zeta$ & $\eta$) obtained from a-Si:H depositions were compared with the model calculations. To further investigate the effect of $t_{ON}$ and $t_{OFF}$ on room temperature electrical conductivity under dark ($\sigma_d$) and under 100 $\frac{mW}{cm^2}$ intensity illumination ($\sigma_{ph}$), aluminum contacts with 0.5 mm spacing were thermally evaporated on the a-Si:H films and current-voltage (I-V) measurements were performed using Keithley 6517B electrometer.

## IV. RESULTS AND DISCUSSION

### A. Comparing Model Calculations with Experiments

For catalytic $SiH_4$ dissociation on filament surface, equations Eq. 1 – Eq. 5 provides the time-dependent concentrations of the surface entities ($M^*$, $SiH_4^*$, $Si^*$ and $H^*$) and the total amount of the generated final products ($N(Si)$ and $N(H_2)$).

#### *1. SiH₄ dissociation under continuous SiH₄ flow*

Fig. 2 shows the calculated $[M^*]$, $[SiH_4^*]$, $[Si^*]$, and $[H^*]$ representing $SiH_4$ dissociation on the filament at temperature $T_f(K)$ under continuous $SiH_4$ flow for model parameters $k_{in} = 10\ s^{-1}$, $k_r = 1.78 \times 10^9 \cdot exp(-\frac{24000}{T_f})\ s^{-1}$, $k_{d1} = 6.01 \times 10^2 \cdot exp(-\frac{9600}{T_f})\ s^{-1}$, $k_{d2} = 3.32 \times 10^3 \cdot exp(-\frac{13200}{T_f})\ s^{-1}$, and $k_{d3} = 2.98 \times 10^3 \cdot exp(-\frac{12000}{T_f})\ s^{-1}$, where time $t_{dep} = 0$ denotes the beginning of precursor flow.

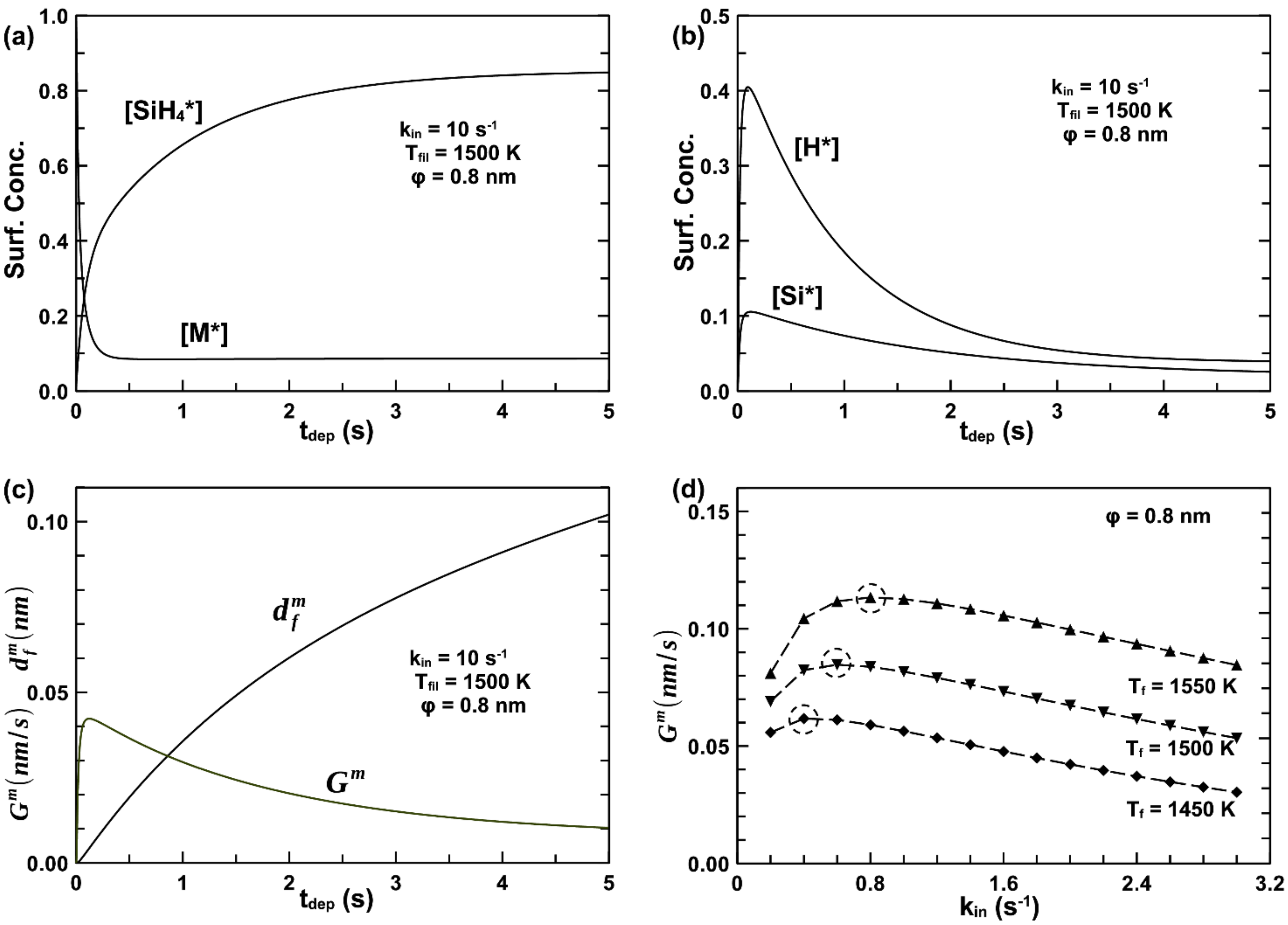

Fig. 2 (a & b) Concentrations $[M^*]$, $[SiH_4^*]$, $[Si^*]$ and $[H^*]$ and (c) the modelled growth rate $G^m$ with corresponding film thickness $d_f^m$ as function of the deposition time $t_{dep}$ for $SiH_4$ dissociation under continuous flow on the 1500 K filament calculated from equations Eq. 1 – Eq. 5. (d) Influence of impingement rate $k_{in}$ on growth rate $G^m$ at $T_f$ = 1450 K, 1500 K and 1550 (encircled data point represents maximum $G^m$).

Starting with a pristine filament ($[M^*] = 1$), the introduction of $SiH_4$(g) at $t_{dep} = 0$ initiates formation of the adsorbed $SiH_4^*$ surface units, that may further desorb back as $SiH_4$(g) or dissociate into $Si^*$ and $H^*$ units in presence of 4 additional $M^*$ sites following reaction R2. The concentration $[SiH_4^*]$ at any instant depends on the adsorption, desorption and dissociation rates at that instance (see Eq. 1). Since $M^*$ sites are in excess in the early stages of precursor exposure, the higher $SiH_4^*$ dissociation rate appears as initial rapid increase in $[Si^*]$, and $[H^*]$. Hence concentrations $[SiH_4^*]$, $[Si^*]$, and $[H^*]$ each are seen to initially increase with $t_{dep}$ in Fig. 2(a, b) accompanied by the decrease in $[M^*]$ according to Eq. 6. Under a continuous precursor flow, although $SiH_4$ adsorption continuous on the accessible $M^*$ sites (see Fig. 2a), beyond a critical time ($t_{dep} \approx 0.12\ s$) the reduced $[M^*]$ slows down $SiH_4^*$ dissociation, leading to gradual decrease in $[Si^*]$ and $[H^*]$ (see Fig. 2b). Analytical expression for the steady state concentrations $[M^*]$, $[SiH_4^*]$, $[Si^*]$, and $[H^*]$ are presented in the the section S1 of the supplementary material.[22]

Since the modelled growth rate $G^m$ is proportional to the rate of $Si(g)$ desorption from the filament surface, which is proportional to $[Si^*]$ according to Eq. 4, $G^m$ too increases with $t_{dep}$ till a maximum value followed by a gradually decrease to a steady state value as shown in Fig. 2 (c) along with the model film thickness $d_f^m$ highlighting the initial non-linear growth regime.

In HWCVD growth experiments, the only accessible parameters related to precursor dissociation are the precursor flow rate (represented here by $k_{in}$) and the filament temperature $T_f$ (that affects $k_r$, $k_{d1}$, $k_{d2}$ and $k_{d3}$). Although the overall $SiH_4$ dissociation kinetics and consequently the steady state growth rates evidently depend upon both $k_{in}$ and $T_f$, this dependence is obscured by the coupled nature of differential equations Eq. 1 – Eq. 5 (see S1 in supplementary material)[22]. Fig. 2 (d) shows the effect of $k_{in}$ on the calculated steady growth rate $G^m$ at different $T_f$ values (assuming $\varphi$ is independent of temperature). At a fixed $T_f$, initially $G^m$ increases with $k_{in}$ until a maximum beyond which it decreases with $k_{in}$. Interestingly with increasing $T_f$, not only is the maximum $G^m$ higher, but this maximum $G^m$ is also obtained for increasingly higher $k_{in}$ value. Encircled data points in Fig. 2 (d), shows that for a 1450 K filament, the maximum $G^m$ (~0.06 nm/s) is obtained for $k_{in} = 0.4\ s^{-1}$, whereas with a 1550 K filament, the maximum $G^m$ (~0.11 nm/s) is obtained for $k_{in} = 0.8\ s^{-1}$.

Experimentally determined growth rate ($G$) for a-Si:H HWCVD deposition for varying $SiH_4$ flow rate ($F_{SiH_4}$) is shown in Fig. 3 at filament temperatures $(T_f)$ 1873 K and 1923 KAs shown here, with increasing $T_f$ from 1873 K to 1923 K, the maximum $G$ increased from 0.67 nm/s (at $F_{SiH_4}$ = 30 sccm) to 1.59 nm/s (at $F_{SiH_4}$ = 40 sccm).

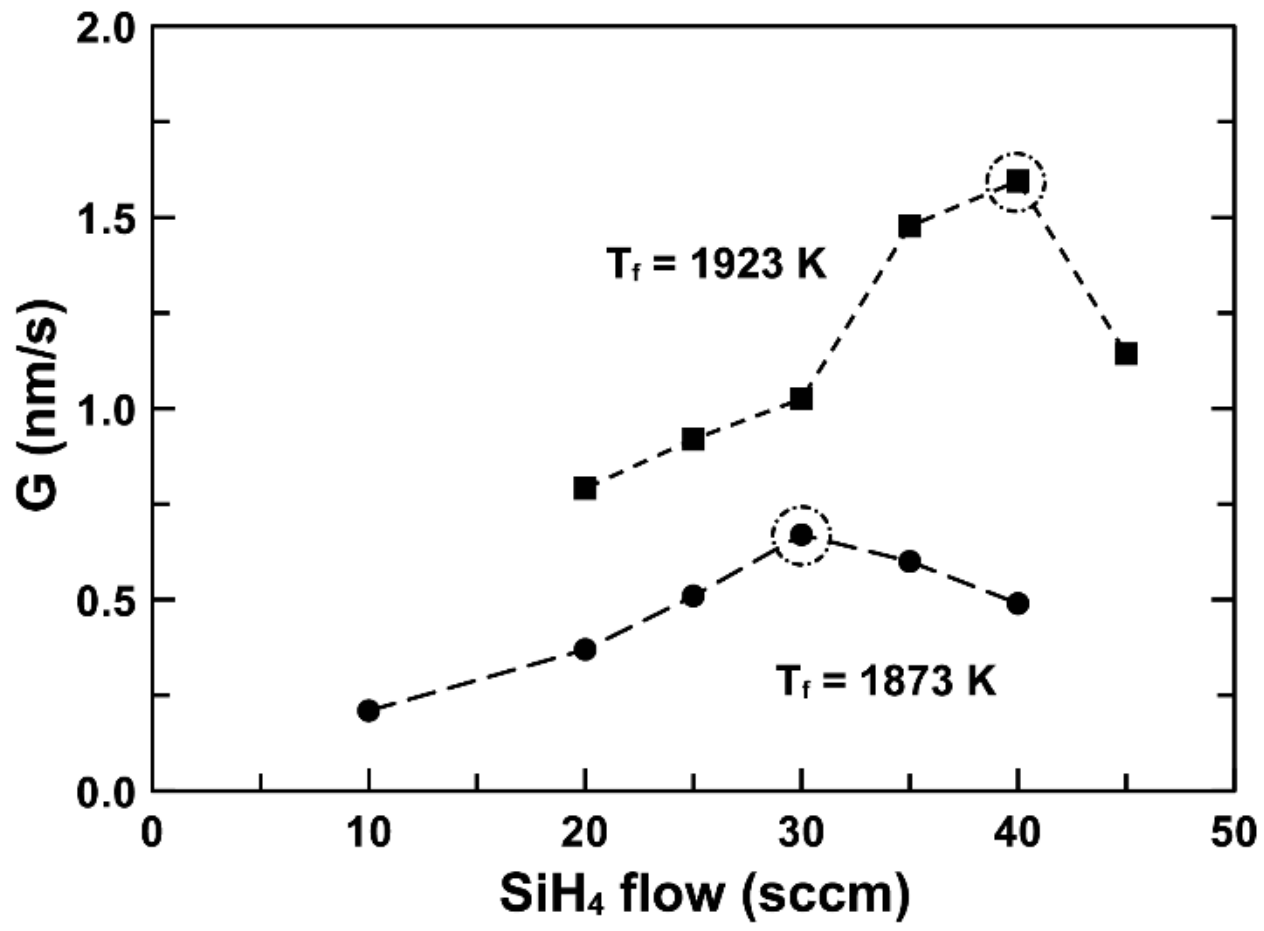

Fig. 3 Effective growth rate $G$ versus $SiH_4$ flow from a-Si:H HWCVD growth experiments for filament temperature $T_f$ 1873 K and 1923 K, respectively. Typical variation in growth rate is within ± 0.04 nm/s of the reported values and the encircled data points represent the maximum growth rate.

Comparison between Fig. 3 and Fig. 2 (d) suggests that our proposed catalytic $SiH_4$ dissociation model was able to describe the effect of precursor flow and filament temperature on HWCVD growth rates to a fair extent. With this qualitative agreement between the calculated and the experimental trends in growth rate, we extended our proposed dissociation model to study a-Si:H growth under pulsed precursor flow.

### *2. $SiH_4$ dissociation under pulsed $SiH_4$ flow*

During pulsed $SiH_4$ flow (See Fig), precursor is introduced periodically over a heated filament for a discrete number of pulses ($N_p$), where the time period ($t_p = t_{ON} + t_{OFF}$) for each pulse comprises of time lengths $t_{ON}$ when $SiH_4$ flow is turned ON and $t_{OFF}$ when the $SiH_4$ flow is turned OFF. Precursor dissociation for a-Si:H growth under pulsed $SiH_4$ flow conditions is modelled with Eqs. (1) – (5) by intermittently setting $k_{in} = 0$ over duration $t_{OFF}$ in every precursor pulse. With this description of the precursor pulse, continuous precursor flow over time $t_{dep}$ could be treated as a limiting case where $N_p = 1$, $t_{ON} = t_{dep}$ and $t_{OFF} = 0$.

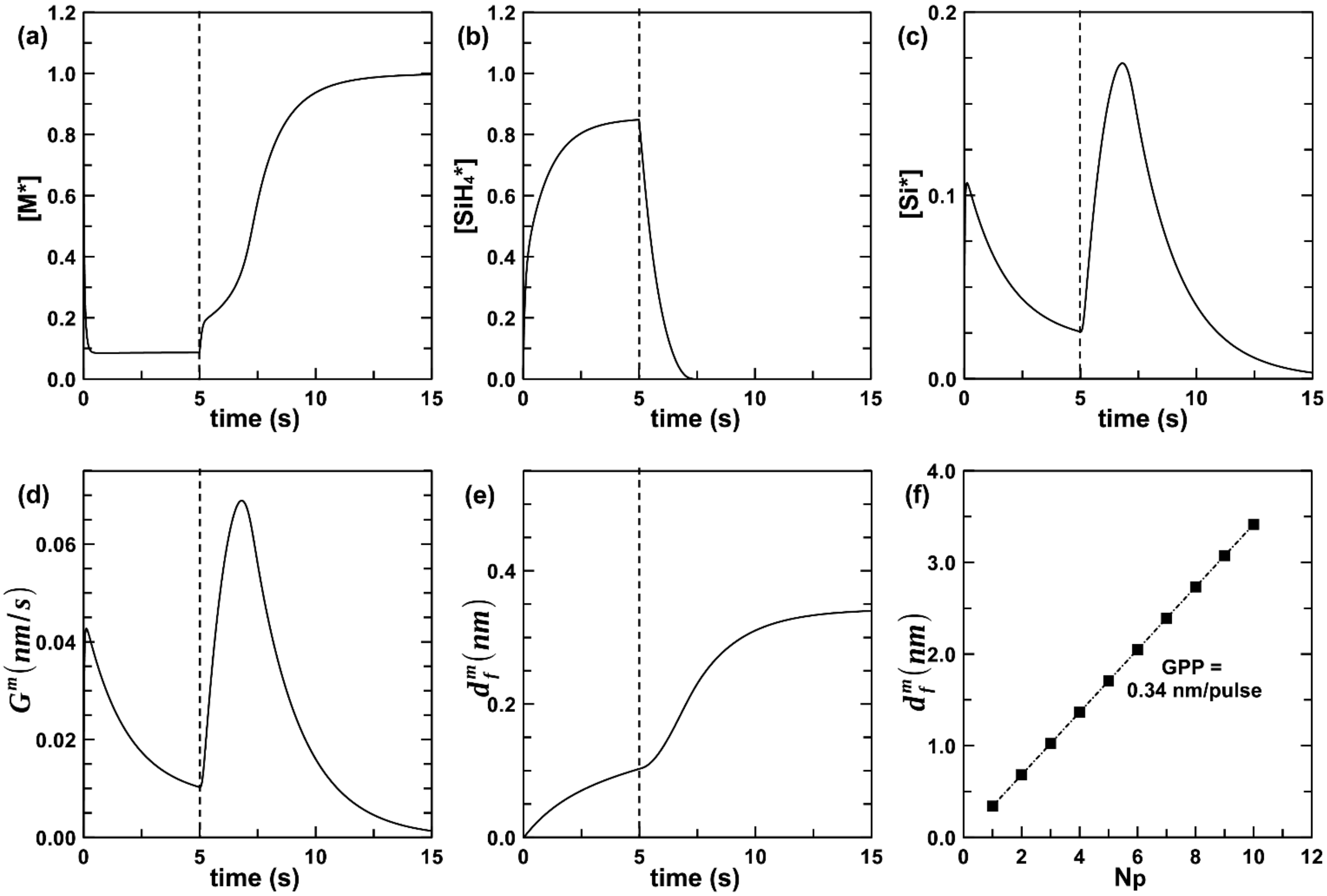

Fig. 4 Evolution of the concentrations (a) $[M^*]$, (b) $[SiH_4^*]$ and (c) $[Si^*]$ calculated for $SiH_4$ dissociation on 1500 K filament surface in the case of pulsed precursor flow with $t_{ON} = 5\ s,\ t_{OFF} = 10s$ and $k_{in} = 10\ s^{-1}$. Also shown here is the (d) instantaneous growth rate $G^m$ (for $\varphi = 0.8$) and (e) the resulting film thickness $d_f^m$. Vertical dotted line at $t = 5s$ represents the moment when $k_{in}$ is set to 0. Although a single pulse is shown in (a) – (e) for clarity, (f) $d_f^m$ increases linearly with number of pulses ($N_p$) with corresponding growth per pulse (GPP) of 0.34 nm/pulse. Calculated $[H^*]$ values are not shown here.

Fig. 4 shows the results from model calculations for precursor dissociation for pulsed the $SiH_4$ flow ($t_{ON} = 5$ & $t_{OFF} = 10s$) over 1500 K filament. Starting with a hot pristine filament ($[M^*] = 1$ at $t = 0$), for time interval $0 < t \leq 5s$ when $k_{in} = 10\ s^{-1}$, the surface concentrations ($[M^*]$, $[SiH_4^*]$, and $[Si^*]$) as well as growth rate $G^m$ in Fig. 4 are seen to be exactly similar to those shown in Fig. 2 (for $t_{dep}\ = 5s$ ). This similarity between Fig. 2 and Fig. 4 is expected since the pristine filament was exposed to 5 s $SiH_4$ flow in both cases. However, for $5s < t \leq 15s$ interval when $k_{in} = 0\ s^{-1}$, only the already existing $SiH_4^*$, $Si^*$ and $H^*$ species can take the surface reaction forward, consequently one notices an interesting behavior in Fig. 4. Under continuous $SiH_4$ flow ($t \leq 5s$), the dissociation rate of $SiH_4^*$ surface species approaches a steady value depending upon the availability of 4 additional $M^*$ sites (See R2). On the contrary, for interrupted $SiH_4$ flow, Fig. 4 (a) —(c) shows $[M^*]$ increases for $t > 5s$, which further facilitates $[SiH_4^*]$ dissociation resulting in higher $[Si^*]$. This change in surface reaction kinetics leads to a higher $G^m$ (Fig. 4 d) and consequently into additional $d_f^m$ (Fig. 4 e) in the $t_{OFF}$ period of the precursor pulsed precursor flow. Introduction of the $t_{OFF}$ between successive precursor pulses affects surface reaction kinetics and consequently the growth ($d_f^m$)-per-pulse (GPP) in two ways: (i) the excess desorption of $Si^*$ from the filament surface because of enhanced $SiH_4^*$ dissociation during $t_{OFF}$, and (ii) partial or complete recovery of $M^*$ on filament surface following desorption of the residual $SiH_4^*$, $Si^*$, and $H^*$ surface species.

Also shown in Fig. 4(f) is the calculated $d_f^m$ values at the end of every precursor pulse, which is seen to increase linearly with the number of pulses ($N_p$) with slope representing growth-per-pulse (GPP) of 0.34 nm/pulse. ~~In order to~~To compare HWCVD growth under continuous and pulsed precursor flow, we define process metrics (i) process throughput given by the ratio $\zeta = \frac{GPP}{(t_{ON}+t_{OFF})}$ that evaluates the effect of $t_{OFF}$ on the overall throughput and (ii) precursor utilization that is proportional to $\eta \propto \frac{GPP}{t_{ON}}$ that represents how efficiently the introduced precursor is utilized towards film growth.

Fig. 5 shows the thickness of ~~HWCVD grown~~HWCVD-grown a-Si:H films ($d_f$) using pulsed $SiH_4$ flow ($F_{SiH_4}$) for a varying number of pulses ($N_p$). Here $F_{SiH_4} = 20\,sccm$ for $t_{ON} = 15\,s$, $F_{SiH_4} = 0\,sccm$ for $t_{OFF} = 45\,s$ and filament temperature at $T_f = 1923 \pm 5\,K$. Under these growth conditions, the measured $d_f$ values were found to linearly increase with $N_p$ with GPP = 28.48 nm/cycle, in agreement with the model thickness trends of Fig. 4(f).

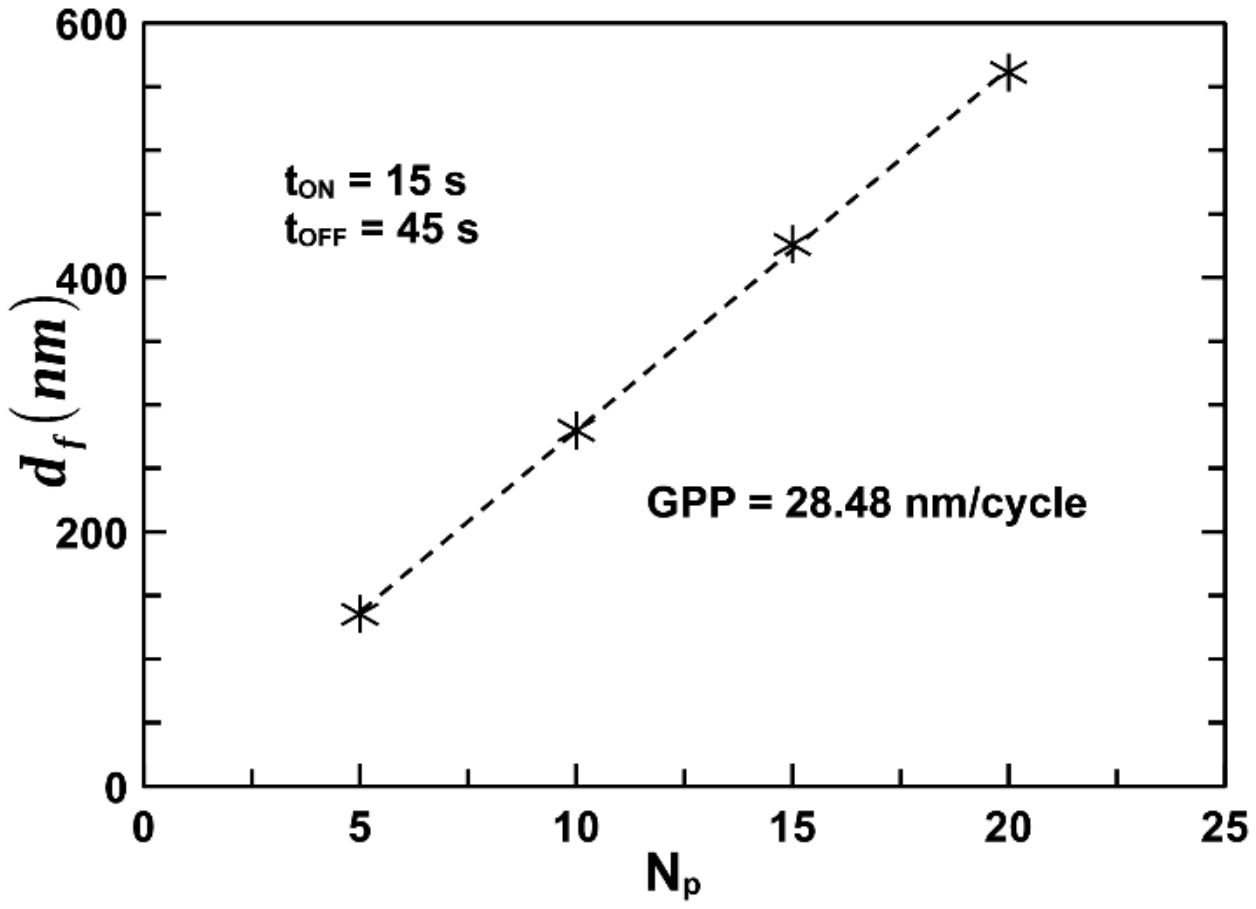

Fig. 5 Film thickness ($d_f$) versus number of pulses ($N_p$) for HWCVD growth of a-Si:H under pulsed $SiH_4$ flow of $F_{SiH_4} = 20\,sccm$ for $t_{ON} = 15\,s$ and $F_{SiH_4} = 0\,sccm$ for $t_{OFF} = 45\,s$ and $T_f = 1923\,K$. The thickness variation was within ±1% of the reported values. The dotted line is the linear fit to thickness data showing a GPP of 28.48 nm//cycle and $R^2 = 0.9997$.

Calculated surface concentrations in Fig. 4 highlights the effect of $t_{ON}$ and $t_{OFF}$ on the resulting GPP and consequently on the throughput and precursor utilization. In Fig. 6(a), the calculated GPP is shown to increase with $t_{ON}$ (for $t_{OFF} = 10\,s$). The film growth during $t_{ON}$ is influenced by the partial or complete re-generation of $M^*$ with residual $[Si^*]$ & $[SiH_4^*]$ on the filament surface over $t_{OFF}$ duration of the preceding precursor pulse. Here the slope $\left(\frac{\Delta GPP}{\Delta t_{ON}}\right)|_{t_{OFF}}$ is seen to decrease with $t_{ON}$, approaching the steady state $G^m$ for a-Si:H growth under continuous precursor flow. The process throughput in Fig. 6(a) was found to go through a maximum at $t_{ON} = 2$. The ratio $\eta$ is seen continuously decreases with $t_{ON}$ (see Fig in supplementary information), thus suggesting precursor pulses with shorter $t_{ON}$ lead to improved precursor utilization.

The effect of varying $t_{OFF}$ on calculated GPP shown in Fig. 6(b) for $t_{ON} = 5s$ follows a remarkably distinct trend as compared to Fig. 6(a). Here $t_{OFF} = 0s$ corresponds to the HWCVD growth under continuous $SiH_4$ flow with GPP equivalent to the thickness $d_f^m$ for $t_{dep} = t_{ON}$ in Fig. 2(c). Considering the catalytic surface reactions (R1-R4), $Si^*$ desorption from filament surface during $t_{OFF}$ interval will result in higher thickness. However, beyond a critical $t_{OFF}$, the $[Si^*]$ would have decreased to an extent (see Fig. 4c) such that $Si^*$ desorption leads to negligible increase in thickness. Hence although the calculated GPP is seen to initially increase with $t_{OFF}$, it was found to approach a self-limiting value of ~0.34 nm/pulse for $t_{OFF} > 6\ s$. ~~Typically~~Typically, longer $t_{OFF}$ (hence longer $t_p = t_{ON} + t_{OFF}$) should lead to longer deposition times and hence poor throughputs. ~~However~~However, the calculated throughputs in Fig. 6(b) shows a steep increase in throughput with $t_{OFF}$ for $t_{OFF} < 2s$, followed by a gradual decrease with longer $t_{OFF}$. For a fixed $t_{ON}$, ratio(?) $\eta$ and GPP follows the same trend with respect to $t_{OFF}$ (See supplementary information for figure), indicating an improved precursor utilization in pulsed deposition ($t_{OFF} > 0$) as compared to continuous deposition.

The effect of $t_{ON}$ and $t_{OFF}$ combination on the GPP and throughput are shown in Fig. 6(c) & Fig. 6(d) respectively for $k_{in} = 10\ s^{-1}$ and $T_f = 1500\ K$. These contours lines are useful in selecting most appropriate precursor pulse for the desired application. For example, $t_{OFF} = 6s$ and $t_{ON} = 7s$ gives highest GPP whereas $t_{OFF} = 2s$ and $t_{ON} = 1s$ would give highest throughput.

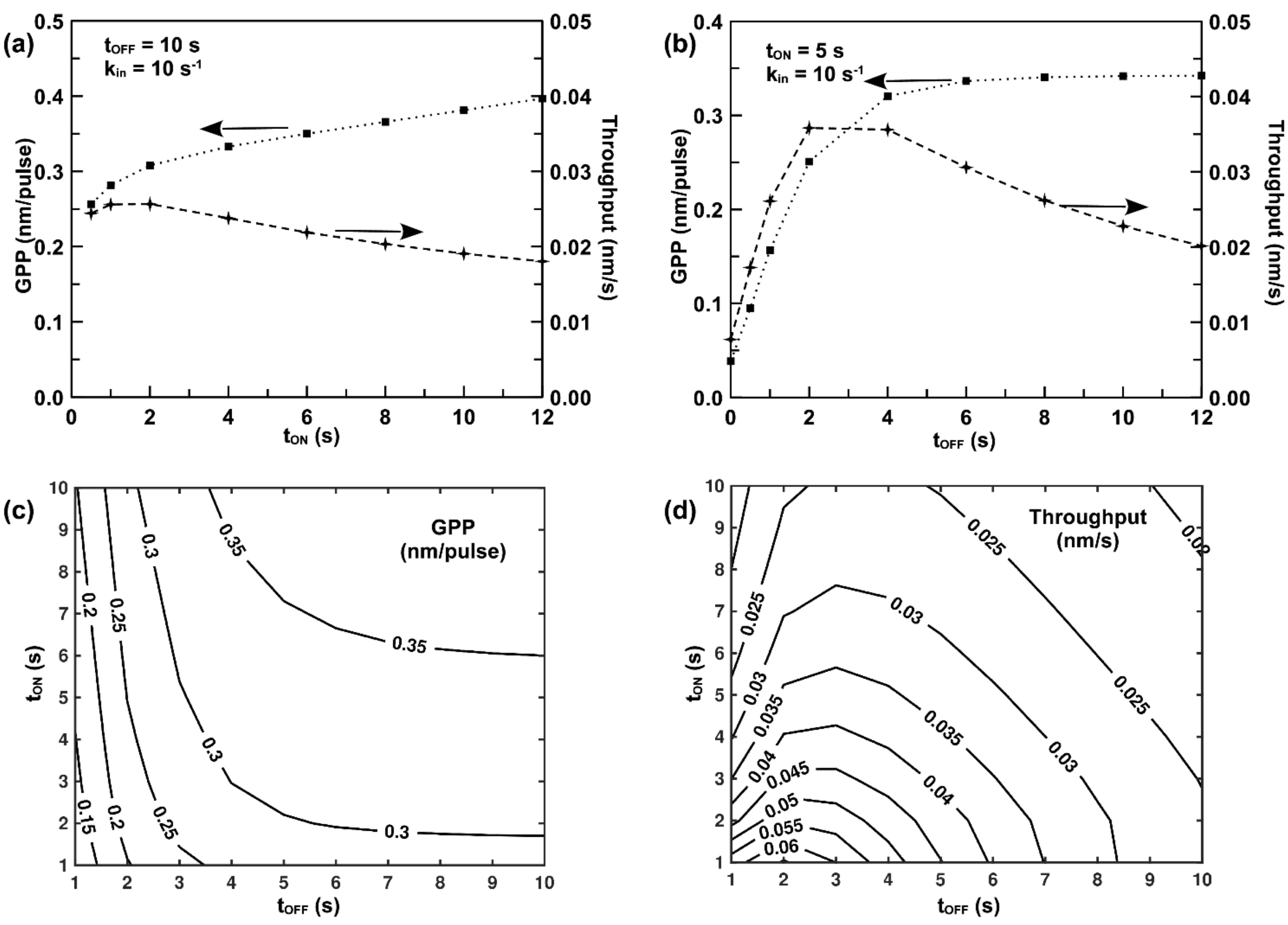

Fig. 6 Effect of (a) $t_{OFF}$ (at $t_{ON} = 5s$) and (b) $t_{ON}$ (at $t_{OFF} = 10s$) on the growth per pulse and effective throughput for modelled a-Si:H HWCVD growth under pulsed $SiH_4$ flow with $k_{in} = 10\ s^{-1}$ and $T_f = 1500\ K$. (c) GPP (nm/pulse) and (d) Process throughput (nm/s) contours illustrating the combined effect of the $t_{OFF}$ & $t_{ON}$ combination.

Model calculations reported in Fig. 6 were performed for a fixed precursor impingement rate $k_{in} = 10\ s^{-1}$ and at a constant filament temperature $T_f = 1500\ K$. Although the influence of $k_{in}$, $k_r$, $k_{d1}$, $k_{d2}$ and $k_{d3}$ on a-Si:H growth under pulsed $SiH_4$ flow has not been extensively investigated in this reported study~~,~~. Table S1 in the supplementary information ~~did demonstrate~~demonstrates a consistent enhancement in the deposited film GPP~~thickness~~ using pulsed precursor flow as compared with continuous precursor flow.

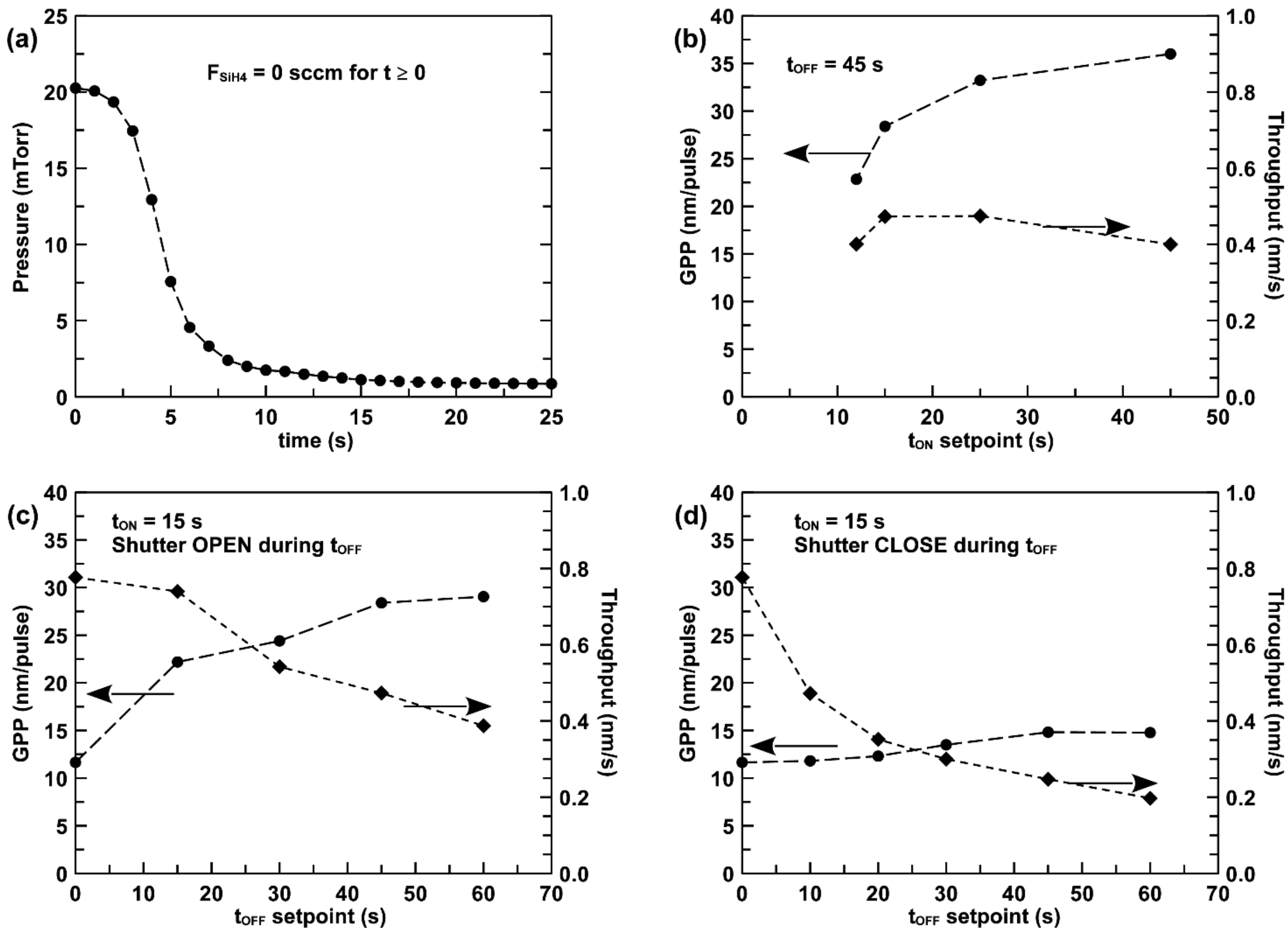

Fig. 7 (a) Reactor pressure vs time after setting $F_{SiH_4} = 0\ sccm$. Dependence of GPP and throughput on (b) $t_{ON}$ (with $t_{OFF} = 45\ s$), (c) $t_{OFF}$ (with shutter OPEN and $t_{ON} = 15\ s$), and (d) $t_{OFF}$ (with shutter CLOSE and $t_{ON} = 15\ s$). For deposition experiments reported in (b) – (d), during $t_{ON}$ duration of the precursor pulse, reactor pressure was 20 mTorr, $SiH_4$ flow rate was $F_{SiH_4} = 20\ sccm$, while filament temperature $T_f = 1923\ K$ over the entire duration. The pulse times ($t_{ON}$ and $t_{OFF}$) shown here are the set points, whereas referring to (a), the actual times are expected to be ($t_{ON} + \delta$) and ($t_{OFF} - \delta$) as described in the text.

Unlike model calculations, in actual experiments, the $SiH_4$ precursor flow $F_{SiH_4}$ over heated filaments cannot be abruptly switched to zero because of the finite response time of mass flow controllers (MFC's) and the residual precursor in the gas-lines connecting MFC and the reactor. This delayed precursor switching is demonstrated by Fig. 7(a) where the measured pressure in the HWCVD reactor is plotted against time for $F_{SiH_4} = 0$ at $t = 0$. Pressure drop represents the evacuation of the residual precursor through the reactor, and as could be seen here this delay ($\delta$) is around ~15 s for near complete evacuation (base pressure of $10^{-5}$ mTorr) after $F_{SiH_4}$ is set to 0 sccm <**do we need to correct capmanometer readings**>. In presence of precursor evacuation delay ($\delta$), the set precursor pulse with durations $t_{ON}$ and $t_{OFF}$ gets modified as precursor pulse with durations $t'_{ON} = t_{ON} + \delta$ and $t'_{OFF} = t_{OFF} - \delta$ respectively. This deviation from the set pulse parameters must be taken into account while analyzing experimentally GPP trends with respect to the set $t_{ON}$ and $t_{OFF}$ values as shown in Fig. 7(b) - Fig. 7(d).

Fig. 7(b) shows the effect of the set $t_{ON}$ (for $t_{OFF} = 45\ s$) on the GPP and process throughput for a-Si:H growth in HWCVD process using pulsed $SiH_4$ flow with filament set to $T_f = 1923\ K$ throughout the deposition. During the precursor flow duration ($t_{ON}$), $F_{SiH_4}$ = 20 sccm and the reactor pressure was 20 mTorr. The experimentally determined GPP and throughput values here followed a similar trend ~~as for~~to that of the calculated GPP versus $t_{ON}$ plot in Fig. 6(a).

The effect of the set $t_{OFF}$ (for $t_{ON} = 15\ s$) on the GPP and process throughput for a-Si:H growth is shown in Fig. 7(c). For $\delta = 15s$, the actual precursor pulse durations were $t'_{ON} = 30s$ and $t'_{OFF} = (t_{OFF} - 15)s$ for each deposition experiment. ~~Consequently~~Consequently, the deposition with $t_{OFF} = 15s$ (or $t'_{OFF} = 0s$) could be considered as a continuous deposition with effective $t'_{ON} = 30s$, thus resulting in GPP (=22.2 nm/pulse), approximately twice that for a continuous deposition with $t_{ON} = 15s$ (11.66 nm/pulse). ~~Moreover~~Moreover, with further increasing $t_{OFF}$, GPP was found to approach a self-limiting value of 29 nm/pulse in agreement with the calculated GPP trend shown in Fig. 6(b). However, contrary to the calculated throughput in Fig. 6(b), the experimental process throughput in Fig. 7(c) decreased with increasing $t_{OFF}$. This difference between calculated and experimental throughput trends may be attributed to the model approximations and/or limited $t_{OFF}$ points considered in experiments.

For a fixed $t'_{ON} = 30s$, GPP trend in Fig. 7(c) could be attributed to $Si^*$ desorption and consequently $M^*$ regeneration on the filament surface in $t'_{OFF}$ interval and enhanced precursor dissociation on regenerated $M^*$ during the $t'_{ON}$ of the following pulse (See Fig S02 in supplementary information). ~~In order to~~To decouple these two contributions~~,~~ and investigate the effect of $M^*$ regeneration alone on GPP, in the following experiments, a shutter was introduced (shutter CLOSE) between the filament and the substrate during the $t_{OFF}$ interval of the precursor pulse to prevent the desorbed $Si^*$ species from reaching the substrate. Fig. 7(d) shows the plot of GPP and process throughput values determined from these experiments versus $t_{OFF}$. For $\delta = 15s$, precursor pulse with $t_{OFF} = 10s$ is equivalent to a continuous deposition and consequently, the GPP values at $t_{OFF}$ = 0s and 10 s are nearly identical. For a fixed $t_{ON} = 15s$ with shutter CLOSE during $t_{OFF}$ interval, GPP increase can only ~~results~~ result from the enhanced precursor dissociation from the recovered $M^*$ concentration during the $t'_{OFF}$ interval of the precursor pulse. Since the $M^*$ recovery is self-limiting in nature (see Fig. 4(a)), the GPP ~~too~~ also approaches a self-limiting value of 14.8 nm/pulse with $t'_{OFF}$ as shown in Fig. 7(d). The process throughput in Fig. 7(d) also continuously decreased with increasing $t_{OFF}$, and as for Fig. 7(c) may be attributed to the model approximations and/or selected experimental conditions. Based on the observed agreement between the experimental and calculated GPP~~C~~ trends in Fig. 6 and Fig. 7, we attempt to investigate the effect of precursor pulsing on the properties of deposited a-Si:H films.

Fig. 8 shows the effect of $t_{OFF}$ on the dark conductivity ($\sigma_d$, measured at 100 V bias) and the ~~photo conductivity~~photoconductivity ($\sigma_{ph}$, measured under 1 Sun illumination and 10 V bias) for a-Si:H films deposited using pulsed $SiH_4$ flow with $t_{ON} = 15\ s$. Compared to a-Si:H films deposited with $t_{OFF} = 0\ s$ (equivalent to continuous $SiH_4$ flow), decrease in $\sigma_d$ and an increase in $\sigma_{ph}$ with increasing $t_{OFF}$ suggests overall improved electrical transport in the a-Si:H films deposited using pulsed $SiH_4$ flow. This could be attributed to a non-uniform growth rate during the precursor pulse (see $G^m$ in Fig. 4(d)) and the film relaxation over $t_{OFF}$ interval. Since this study was restricted to precursor dissociation modelling and its effect on growth characteristics, detailed structural and electronic characterization of a-Si:H films deposited under pulsed precursor flow is beyond its scope. Also, in this study, we did not consider the effect of desorbing $H^*$ species (see reaction R4) on GPP and structural/electronic properties of the resulting film. The effect of $H/H_2$ on the growth and characteristics of a-Si:H films will be reported in ~~the~~ future work.

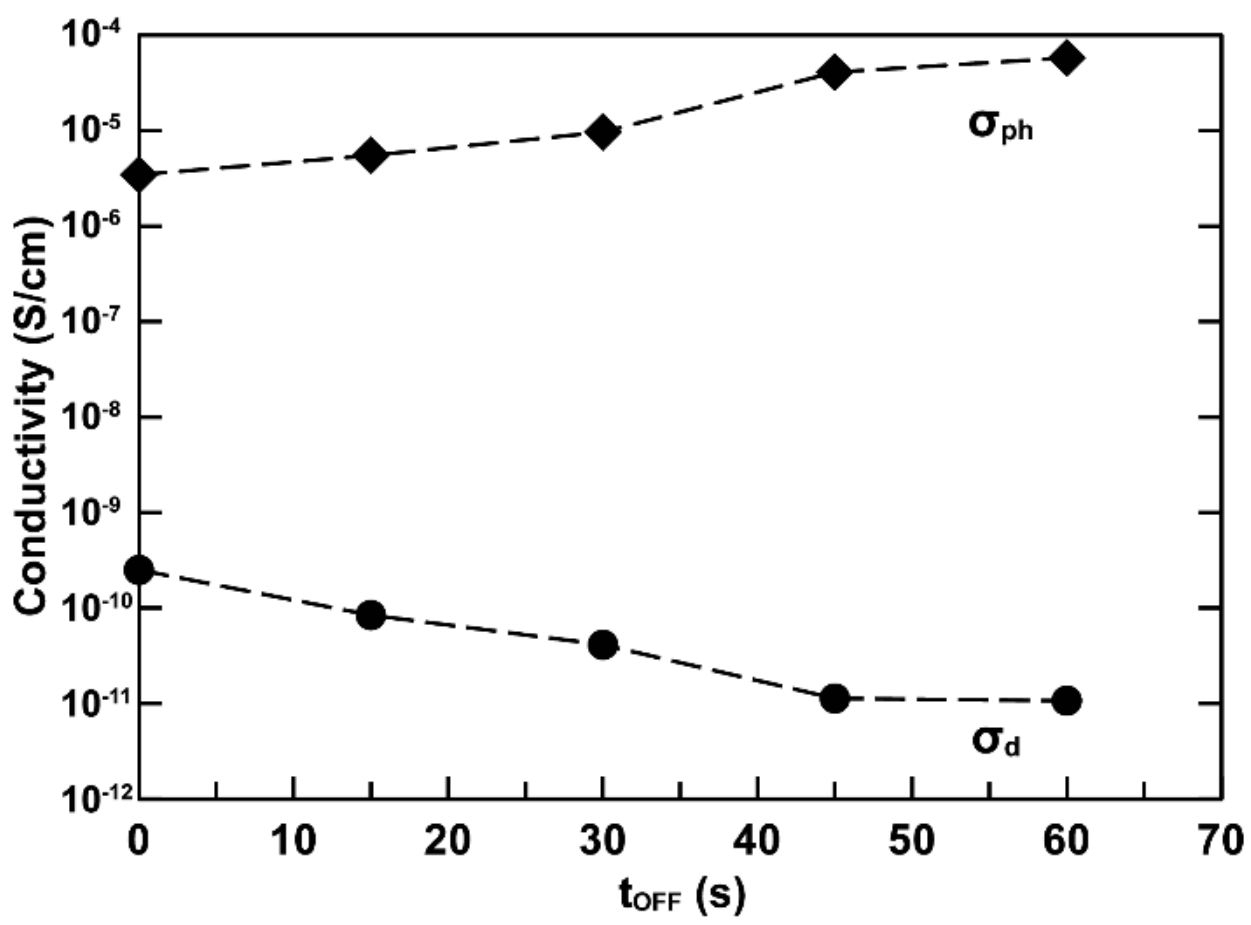

Fig. 8 Effect of $t_{OFF}$ on dark conductivity ($\sigma_d$) and photoconductivity ($\sigma_{ph}$) measured on a-Si:H films deposited using pulsed $SiH_4$ flow with $t_{ON} = 15\ s$. For these films $T_f = 1923\ K$ and shutter was OPEN throughout deposition, and during $t_{ON}$ reactor pressure was 20 mTorr, $SiH_4$ flow rate was $F_{SiH_4} = 20\ sccm$.

### V. Summary and Conclusion

In this study, we proposed a model for catalytic dissociation of $SiH_4$ in HWCVD growth of a-Si:H films based on the surface reactions R1 – R4. Calculated growth rate ($G^m$) versus precursor flow ($k_{in}$) trends were found to agree with the corresponding experimentally determined trends. Hence the proposed dissociation model was extended to a-Si:H growth under pulsed $SiH_4$ flow represented by durations $t_{ON}$ and $t_{OFF}$. Growth-per-pulse (GPP) dependence on $t_{ON}$ and $t_{OFF}$ determined from a-Si:H growth experiments with pulsed $SiH_4$ flow were seen to qualitatively agree with the calculated GPP trends. Considering a reactor evaluation delay $\delta \approx 15s$ during a-Si:H growth using a fixed $t_{ON} = 15s$, ~30 % increase in GPP with $t_{OFF}$ (from 22.2 nm/pulse at $t_{OFF} = 15s$ to 29.6 nm/pulse at $t_{OFF} = 60s$) demonstrates improved utilization of the introduced $SiH_4$ precursor. Moreover, the dark conductivity ($\sigma_d$) and the photoconductivity ($\sigma_{ph}$) measurements suggested improved electronic transport in a-Si:H films deposited with pulsed $SiH_4$ flow ($\sigma_d = 1.1 \times 10^{-11}\ S/cm$ and $\sigma_{ph} = 5.8 \times 10^{-5}\ S/cm$ for $t_{OFF} = 60s$) than with continuous deposition ($\sigma_d = 2.5 \times 10^{-10}\ S/cm$ and $\sigma_{ph} = 3.5 \times 10^{-6}\ S/cm$ for $t_{OFF} = 0s$).

The agreement between our modelling and experimental a-Si:H growth data indicates that the catalytic dissociation mechanism significantly contributes towards the overall precursor dissociation over heated Ta filaments in HWCVD growth. We have shown that $SiH_4$ utilization in HWCVD growth may be improved by adopting a pulsed precursor flow process. Although not included here, the improved HWCVD process model incorporating silicide formation over heated Ta filament and the effect of $H/H_2$ on growth and properties of a-Si:H films will be reported in the future.

<add write ups for all sections in the supplementary data file>

## VI. SUPPLEMENTAL MATERIAL

Text material that may not be of interest to all readers, long data tables, multimedia, and computer programs may be deposited as supplementary materials. An article can have only one reference citing the supplemental material within the article. All citations of the supplemental material in the text must link to that reference. Information about depositing supplemental material may be found in Supporting Data in our Author Resource Center.

## ACKNOWLEDGMENTS

Typically, standard acknowledgments include financial support and technical assistance, and may include dedications, memorials, and awards. Check with the Editorial Office for suitability of an acknowledgment if there is any question. To indicate the author, use initials. For example, “B.A. wishes to thank A. Loudon for technical assistance. C.A. wishes to thank Anytown University for use of their equipment.”

Note: the Acknowledgment section is not a numbered section.

## REFERENCES

[1] A. Pandey, S. Bhattacharya, S. Alam, S. Manna, S. Sadhukhan, S.P. Singh, and V.K. Komarala, "Influence of Very High-Frequency PECVD Hydrogen Plasma Treatment on Intrinsic Amorphous Silicon Passivation Stack: Impact on Silicon Heterojunction Solar Cell Performance," ACS Appl. Energy Mater. **8**(1), 366–375 (2025).
[2] S. Jiang, C. Li, J. Du, D. Wang, H. Ma, J. Yu, and A. Nathan, "Thin-Film Transistor Digital Microfluidics Circuit Design with Capacitance-Based Droplet Sensing," Sensors **24**(15), 4789 (2024).
[3] M. Agarwal, A. Munjal, and R. Dusane, "To Demonstrate the Potential Application of 'Low Temperature and High Performance Silicon Heterojunction Solar Cells Fabricated Using HWCVD' in Wireless Sensor Network: An Initial Research," Journal of Solar Energy Engineering **140**(4), 041002 (2018).
[4] K. Nepal, A. Gautam, C. Ugwumadu, and D.A. Drabold, "New models of clean and hydrogenated amorphous silicon surfaces," Journal of Non-Crystalline Solids **660**, 123517 (2025).
[5] R.O. Dusane, "HWCVD: A Potential Tool for Silicon-Based Thin Films and Nanostructures," in *Recent Advances in Thin Films*, edited by S. Kumar and D.K. Aswal, (Springer Singapore, Singapore, 2020), pp. 455–478.
[6] H. Matsumura, "Current Status of Catalytic Chemical Vapor Deposition Technology ━ History of Research and Current Status of Industrial Implementation ━," Thin Solid Films **679**, 42–48 (2019).
[7] W. Zheng, and A. Gallagher, "Radical species involved in hotwire (catalytic) deposition of hydrogenated amorphous silicon," Thin Solid Films **516**(6), 929–939 (2008).
[8] H.L. Duan, G.A. Zaharias, and S.F. Bent, "The effect of filament temperature on the gaseous radicals in the hot wire decomposition of silane," Thin Solid Films **395**(1–2), 36–41 (2001).
[9] S.K. Soni, A. Phatak, and R.O. Dusane, "High deposition rate device quality a-Si:H films at low substrate temperature by HWCVD technique," Solar Energy Materials and Solar Cells **94**(9), 1512–1515 (2010).
[10] R.E.I. Schropp, "Industrialization of Hot Wire Chemical Vapor Deposition for thin film applications," Thin Solid Films **595**, 272–283 (2015).
[11] A. Pflüger, C. Mukherjee, and B. Schröder, "Optimization of process parameters in a large-area hot-wire CVD reactor for the deposition of amorphous silicon (a-Si:H) for solar cell application with highly uniform material quality," Solar Energy Materials and Solar Cells **73**(3), 321–337 (2002).
[12] H. Meng, X. Wu, G. Zhong, and L. Zhou, "Optimization of HWCVD-grown i-a-Si:H films through synergistic adjustment of hydrogen dilution ratio and hot-wire temperature," Applied Physics Letters **126**(19), 193904 (2025).
[13] H. Matsumura, H. Umemoto, and A. Masuda, "Cat-CVD (hot-wire CVD): how different from PECVD in preparing amorphous silicon," Journal of Non-Crystalline Solids **338–340**, 19–26 (2004).
[14] P.A. Frigeri, O. Nos, and J. Bertomeu, "Degradation of thin tungsten filaments at high temperature in HWCVD," Thin Solid Films **575**, 34–37 (2015).
[15] S. Morrison, and A. Madan, "Deposition of amorphous and microcrystalline silicon using a graphite filament in the hot wire chemical vapor deposition technique," Journal of Vacuum Science & Technology A: Vacuum, Surfaces, and Films **19**(6), 2817–2819 (2001).
[16] O. Nos, P.A. Frigeri, and J. Bertomeu, "Technological solution for the automatic replacement of the catalytic filaments in HWCVD," Thin Solid Films **575**, 30–33 (2015).
[17] T. Muneshwar, and K. Cadien, "*AxBAxB* … pulsed atomic layer deposition: Numerical growth model and experiments," Journal of Applied Physics **119**(8), 085306 (2016).
[18] J.K. Holt, M. Swiatek, D.G. Goodwin, and H.A. Atwater, "The aging of tungsten filaments and its effect on wire surface kinetics in hot-wire chemical vapor deposition," Journal of Applied Physics **92**(8), 4803–4808 (2002).
[19] C.E. Sveen, and Y. Shi, "Effect of filament temperature and deposition time on the formation of tungsten silicide with silane," Thin Solid Films **519**(14), 4447–4450 (2011).
[20] K. Honda, K. Ohdaira, and H. Matsumura, "Study of Silicidation Process of Tungsten Catalyzer during Silicon Film Deposition in Catalytic Chemical Vapor Deposition," Jjap **47**(5R), 3692 (2008).
[21] M. Agarwal, A. Pawar, N. Wadibhasme, and R. Dusane, "Controlling the c-Si/a-Si:H interface in silicon heterojunction solar cells fabricated by HWCVD," Solar Energy **144**, 417–423 (2017).
[22]"Supplementary material for pulsed HWCVD," (n.d.).